\documentclass[useAMS,usegraphicx,usenatbib]{mn2e}

\usepackage{amssymb}
\usepackage{times}
\usepackage{epsfig}
\usepackage{color}

\def\Rdisc{R_{\rm disc}}
\def\Rbb{R_{\rm bb}}
\def\Lx{L_{\rm X}}
\def\Fx{F_{\rm X}}
\def\Tdisc{T_{\rm disc}}
\def\Kdisc{K_{\rm disc}}
\def\Tbb{T_{\rm bb}}
\def\Kbb{K_{\rm bb}}
\def\Tpfree{T_{\rm pfree}}
\def\Kpfree{K_{\rm pfree}}
\def\NH{N_{\rm H}}
\def\Ebr{E_{\rm br}}
\def\Ecut{E_{\rm cut}}

\def\Msun{{\rm M}_{\odot}}

\def\chired{\chi^2_{\rm red}}
\def\ergs{{\rm erg\,s^{-1}}}
\def\ergcm2s{{\rm erg\,cm^{-2}\,s^{-1}}}
\def\cm2{{\rm cm^{-2}}}

\def\swift{{\it Swift}}
\def\xmm{{\it XMM-Newton}}
\def\chandra{{\it Chandra}}

\def\Ho{Ho II X-1}

 \def\aj{AJ}%
\def\apj{ApJ}%
\def\apjl{ApJ}%
\def\aap{A\&A}%
\def\mnras{MNRAS}%
\def\pasj{PASJ}%
\def\apjs{ApJS}%
\def\nat{Nat}
\def\apss{Ap\&SS}
\title[Spectral curvature in Holmberg II X-1]{Evolution of the spectral curvature in the ULX Holmberg II X-1}
\author[Jari J. E. Kajava et al.]{J.\,J.\,E.\,Kajava,$^{1}$\thanks{E-mail: jari.kajava@oulu.fi, juri.poutanen@oulu.fi} J. Poutanen$^1$\footnotemark[1], S.\,A. Farrell,$^{2,\,3}$ F. Gris{\'e}$^{4}$ and P. Kaaret$^{4}$  \\
$^1$Astronomy Division, Department of Physics, P.O.Box 3000, 90014 University of Oulu, Finland \\
$^2$Sydney Institute for Astronomy (SIfA), School of Physics, The University of Sydney, NSW 2006, Australia\\
$^3$Department of Physics and Astronomy, University of Leicester, University Road, LE1 7RH, Leicester, UK\\
$^4$Department of Physics and Astronomy, University of Iowa, Val Allen Hall, Iowa City, IA 52242, USA \\
}

\begin{document}


\pagerange{\pageref{firstpage}--\pageref{lastpage}} \pubyear{2012}

\maketitle

\label{firstpage}

\begin{abstract}
Ultraluminous X-ray sources (ULXs) are interesting systems as they can host intermediate-mass black holes. 
Alternatively, ULXs can represent stellar-mass black holes accreting at super-Eddington rates. 
Recently spectral curvature or breaks at energies above a few keV have been detected in high quality ULX spectra.
These spectral features have been taken as evidence against the intermediate-mass black hole case.
In this paper, we report on a new \xmm\ observation of the ULX Holmberg II X-1 that also shows a clear spectral break at approximately 4 keV. 
This observation was performed during a low luminosity state of the system and by comparing this new data to a high luminosity state \xmm\ observation, we can conclude that the spectral break energy increases with luminosity.
This behaviour is different to a ULX in the Holmberg IX galaxy, where an opposite trend between the luminosity and the spectral break energy has been claimed.
We discuss mechanisms that could explain this complex behaviour.

\end{abstract}


\section{Introduction}\label{sec:intro}

Ultraluminous X-ray sources (ULXs) -- detected in many nearby galaxies -- are defined as (non-nuclear) systems with X-ray luminosities exceeding $\sim\!10^{39}\,\ergs$ (see \citealt{FS11}, for a recent review). 
ULXs are fascinating sources for several reasons.
One popular interpretation is that they host so called intermediate-mass black holes (IMBH), with masses in the range of $\sim\!100\!-\!10^5\,\Msun$ \citep{CM99}.
Thus, they could ``bridge the gap" between the $\sim \! 10\, \Msun$ stellar mass black holes (StMBH) seen in the Milky way \citep{MM06} and the super-massive black holes (SMBH; $M\!\sim\!10^6\!-\!10^{10}\,\Msun$) in the centres of galaxies (e.g. \citealt{FF05}).
Potentially, therefore, ULXs could help us understand how SMBHs have grown to have their large masses \citep{EMT01}.
The most luminous ULXs -- like ESO 243--49 HLX-1 ($\Lx \! \sim \! 10^{42}\,\ergs$, \citealt{FWB09}) and ULX M82 X-1 ($\Lx\!\sim\!10^{41}\,\ergs$, \citealt{KPZ01}) -- are also the best IMBH candidates, 
as they show similar spectral components and spectral states as Galactic StMBHs \citep{GBW09,FK10,SFL11}, but with luminosities and spectral properties consistent with an IMBH. 

M82 X-1 and ESO 243--49 HLX-1 both reside near (or within) young star clusters (see \citealt{PBH04,KYH07,FSP11}).
These sites are ideal for an IMBH to form and for it to capture a stellar companion from which to accrete \citep{PBH04}.
However, the devil is in the detail.
Recent simulations show that IMBHs are bound to these clusters, whereas a large fraction of StMBHs are ejected from them \citep{MRZ11}.
Because many ULXs tend to be located near, but not within, these young clusters (see e.g. \citealt{KAG04} and Poutanen et al. 2012, in preparation) 
this suggests that a majority of ULXs are not IMBHs.
Furthermore, several studies now clearly indicate that the majority of ULXs are connected to recent star formation episodes \citep{GGS03,SGT04,MGS12} 
and that ULXs are located near star formation regions within the host galaxies \citep{STS09}.
A natural explanation to these findings -- and to the high luminosities of ULXs -- is very high (super-Eddington) accretion rate onto StMBHs in high-mass X-ray binaries (\citealt{GGS03,K04,PLF07,MGS12}).
Such an accretion regime is very challenging from a theoretical point of view, 
because at super-Eddington accretion rates many effects -- such as winds/outflows and advection -- can strongly affect the accretion flow properties (e.g. \citealt{PLF07}).
Therefore, the spectrum of a super-Eddington disc can be very complex, 
and it is difficult to compare observational X-ray data of ULXs to these models as they usually lack clear identifiable predictions.

Even though a majority of ULXs could be super-Eddington StMBHs in high-mass X-ray binaries, some can still be IMBHs, and population studies do not provide the means to differentiate between different alternatives for a given ULX.
In the absence of black hole mass measurements through radial velocity studies of optical counterparts of ULXs (see \citealt{RGG11} for such an attempt), 
indirect black hole mass proxies must be used instead.
Many of these attempts during the last decade are based on X-ray spectroscopic measurements using \xmm\ or \chandra\ observations (see e.g. \citealt{MFM03}).
Two common spectral features in ULXs (with respect to simple power-law fits) are typically used: 
the soft excess below 2 keV, and spectral curvature/breaks above 2 keV (see e.g. \citealt{SRW06,GRD09} and the review of \citealt{FS11}).
The nature of the soft excess has been a matter of debate. 
It can be related to direct accretion disc emission (IMBH case, \citealt{KCP03,MFM03,SFL11,FSP11}), 
but for several ULXs the soft excess temperature does not follow the expected luminosity scaling $L\!\propto\!T^4$ (\citealt{FK07,S2007,KP09}).
In these cases the soft excess can instead be associated with an outflow \citep{PLF07,KP09} or it can be emission from a cold outer disc around StMBHs \citep{GRD09}.
It is also possible that in some cases the soft excess is an artefact of ionized absorption below 2 keV and incorrect modelling of the continuum emission \citep{GS06}.

On the other hand, the spectral curvature/breaks are less debated and controversial as regards to the IMBH vs. StMBH debate.
These features can be associated with high optical depth, low temperature coronae \citep{SRW06,GRD09}.
The presence of these breaks in ULX spectra and the absence of them in StMBH or SMBH spectra supports the fact that these ULXs are in a different accretion state. 
This thus fits in better with the idea that the bulk of ULXs are super-Eddington StMBHs rather than sub-Eddington IMBHs.
However, only a small fraction of ULX observations have sufficient data quality at the 5--10 keV range to see these breaks.
Therefore, additional high quality X-ray data are needed to study them, as these features could provide a method of distinguishing super-Eddington StMBHs from IMBH candidates.
In this paper, we report on a new \xmm\ observation of bright and well-studied ULX Holmberg II X-1 that shows this spectral feature.

\section{Target and observations}\label{sec:observations}

The irregular dwarf galaxy Holmberg II at a distance of $3.39$ Mpc (\citealt{KDG02}) hosts an ULX (hereafter \Ho) that has been studied extensively in recent years.
\citet{MLH01} found in their joint {\it ASCA} and {\it ROSAT} spectral modelling that several spectral components are needed to explain the data.
During \xmm\ monitoring in 2002, the system was seen in a peculiar ``low/soft'' state \citep{DMG04}, which is also seen in NGC 5204 ULX and NGC 1313 ULX-1 \citep{KP09}.
In 2004, a 100ks \xmm\ observation of this ULX was performed.
Based on these data \citet{GRR06} argued that the lack of short term variability is inconsistent with the IMBH scenario (see \citealt{HVR09}, for comparison with other ULXs).
This 2004 dataset has been extensively used in recent years \citep{SRW06,GS06,FK09,GRD09,KP09}, 
and it is one of the few ULX observations that show a spectral break at $\sim\!5.4$ keV \citep{SRW06,GRD09}.
\Ho\ has an extended emission line nebula around it, that is powered by photo-ionization \citep{PM02}.
The line emission diagnostics require an X-ray photo-ionization source to have a luminosity of the order of $10^{40}\,\ergs$, which suggest that the X-ray emission of \Ho\ cannot be strongly beamed \citep{KWZ04}.
This is further strengthened by infrared line diagnostics \citep{BDW10} and radio observations \citep{MMN05}.
Thus, we are quite confident that the X-ray emission in \Ho\ is roughly isotropic.

\begin{figure}
\centerline{\epsfig{file=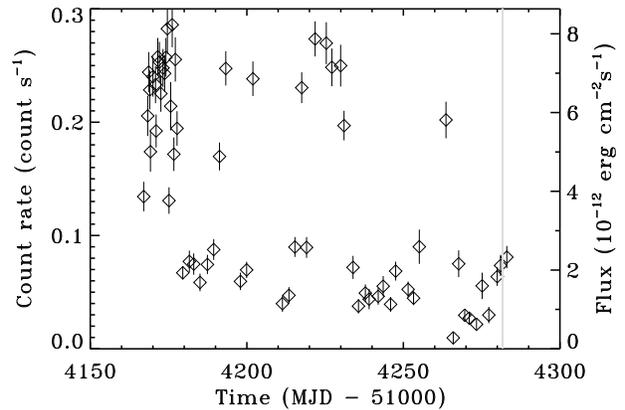,width=\linewidth}}
\caption{
The light curve from the \swift/XRT monitoring campaign in the 0.3--10 keV band \citep[see][]{GKF10}. 
The flux scaling in the second Y-axis was obtained from the {\sc webpimms} tool, using values $\Gamma\!=\!2.7$ and $\NH\!=\!1.5\!\times\!10^{21}\,\cm2$. 
The grey vertical strip indicates the time of the \xmm\ observation.}
\label{fig:xrt_lc}
\end{figure}

Recently, \Ho\ was monitored with \swift\ X-ray telescope (XRT) for four months and the X-ray flux was seen to fluctuate by up to a factor of $\sim$14 during these observations (see \citealt{GKF10} and Fig. \ref{fig:xrt_lc}).
The lowest flux values seen $\Fx\!\sim\!5.8\!\times\!10^{-13}\,\ergcm2s$ indicated that the luminosity dropped down to $\Lx\!\sim\!8\!\times\!10^{38}\,\ergs$ 
(this estimation was made using the {\sc webpimms} tool for the parameters: \swift/XRT count rate 0.02 count s$^{-1}$, photon index $\Gamma\!=\!2.7$ and the hydrogen column density $\NH\!=\!1.5\!\times\!10^{21}$ cm$^{-2}$, see \citealt{GKF10}).
Following the drop into a low luminosity state, 
we triggered an \xmm\ target of opportunity (ToO) observation on 2010 March 26 (MJD 55281, OBSID 0561580401), 
because this luminosity level is comparable to the brightest Galactic StMBHs.
At such low luminosity regime, we could directly compare the spectral properties between this ULX and the Galactic StMBHs, and possibly see if they are similar.
However, as Fig. \ref{fig:xrt_lc} indicates, the initial goal of our observation was not reached because the source underwent a minor flare prior to the observation.
Fortunately, however, this long \xmm\ observation provided the best quality spectrum of this ULX in its peculiar ``low/soft" state \citep{DMG04}.
We also re-analysed the long 2004 April 15 observation (MJD 53110, OBSID 0200470101) as a reference for the higher luminosity state.

\begin{figure}
\centerline{\epsfig{file=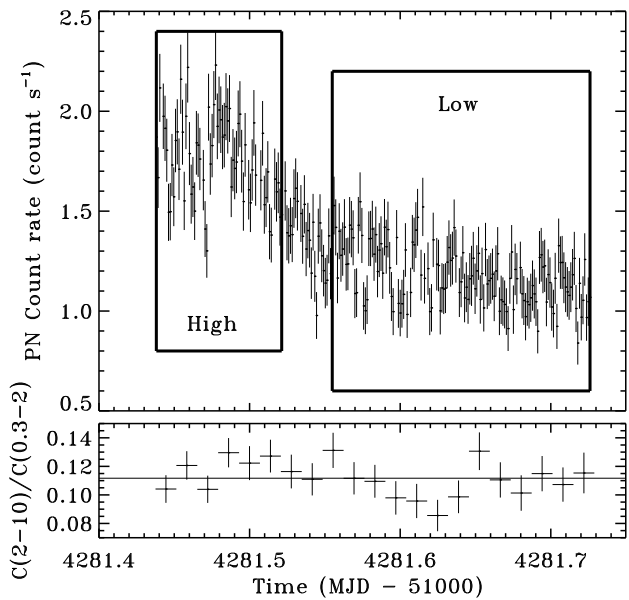,width=\linewidth}}
\caption{
Top panel: EPIC-pn light curve in the 0.3--10 keV band. 
The light curve is background subtracted and corrected using the {\sc epiclccorr} tool. 
Bottom panel: the hardness ratio (2--10 keV band count rate divided by 0.3--2 keV count rate), which does not vary as the flux decays by $\sim$50 per cent.}
\label{fig:pn_lc_hr}
\end{figure}

We processed the Observation Data Files (ODF) with the XMM-SAS version 11 and using methods recommended in XMM-SAS data analysis threads.\footnote{http://xmm.esac.esa.int/sas/current/documentation/threads/}
We used the {\sc epproc} and {\sc emproc} pipelines to produce the calibrated event files for EPIC-pn and EPIC-mos instruments, respectively.
The event selection was done using flag$=$0 and pattern $\le$4 for EPIC-pn and \#XMMEA\_EM and pattern $\le$12 for EPIC-mos.
We then produced light curves in the 10--12 keV band for each instrument over the entire detector to look for background flares.
For the 2004 data we included only the time periods where the count rates were below the recommended $0.4$ count s$^{-1}$ for EPIC-pn and $0.35$ count s$^{-1}$ for EPIC-mos instruments.
For the 2010 data, we found that only the latter part of the observation was affected by background flaring. 
We ignored this part of the observation for all the instruments, even though in few occasions the background count rate dropped below $0.35$ count s$^{-1}$ for EPIC-mos. 
The resulting useful exposure times for the 2010 data were 20ks for EPIC-pn and 25ks for EPIC-mos1 and mos2.
Similarly, for the 2004 dataset, the good exposure times were 35ks, 55ks and 56ks for EPIC-pn, mos1 and mos2, respectively.
The spectrum was very soft during both observations (photon indices $\Gamma\!\approx\!2.7$, see below) and thus the count rate in the 5--10 keV spectral band was very low.
As the spectral shape above $\sim\!5$ keV is of significant importance to study the spectral breaks (\citealt{SRW06,GRD09}),
we determined optimal source extraction radii by generating images for all the EPIC cameras in the 5--10 keV band (from the calibrated and background flare filtered event files) 
and then using the {\sc eregionanalyse} tool to maximize the signal to noise (S/N) ratio in this band.
For the 2004 data, this procedure resulted in source extraction radii of 40\arcsec, 37\arcsec\ and 38\arcsec\ for EPIC-pn, mos1 and mos2, respectively, 
and for the 2010 data 27\arcsec, 22\arcsec\ and 23\arcsec.
We used a background region of 40\arcsec\ radius from a nearby source-free region that was located in the same EPIC-pn CCD chip as the source.
We then used the {\sc especget} tool to extract the spectral data and response files.
The spectra were binned using the {\sc grppha} tool to have a minimum of 20 counts per spectral bin.

\section{Spectral analysis}\label{sec:spectra}

We fitted the spectral data in {\sc xspec v.12.6} using powerlaw based models and phenomenological thermal models, which are typically used for modelling ULX spectra.
We tabulated these best fitting spectral parameters to Table \ref{tab:bestfits} and the quoted errors correspond to the 90 per cent confidence level for a single parameter.
As many of the spectral parameters in these models are similar, we first introduce them here.
We used a {\sc constant} parameter to account for possible cross-calibration uncertainties between the three EPIC instruments.
We fixed this parameter to unity for EPIC-pn, and allowed mos1 and mos2 normalizations to vary.
For the 2004 data, the {\sc constant} parameter values were $\approx\!1.07$ and $\approx\!1.08$ for EPIC-mos1 and mos2, respectively, 
and for the 2010 data they were $\approx\!1.05$ and $\approx\!1.03$, independent of the spectral models used.
We used the {\sc tbabs} model to account for the interstellar absorption using \citet{WAM2000} abundances, parametrized by the column density $\NH$.
The {\sc powerlaw} model is determined by the photon index $\Gamma$ and its normalization $K_{\rm \Gamma}$.
The cutoff power-law model {\sc cutoffpl} has one additional parameter with respect to the {\sc powerlaw} model: 
the e-folding energy $\Ecut$ (in keV), so that the photon flux $N(E)\!\propto\!E^{-\Gamma}\exp(-E/\Ecut)$.
The broken power-law model {\sc bknpower} has two additional parameters: the break energy $\Ebr$ (in keV) and $\Gamma_2$ ($\Gamma_1 \equiv \Gamma$). 
Black body model {\sc bbodyrad} (hereafter {\sc bb}) is parametrized by a temperature $\Tbb$ and normalization $\Kbb\!=\!(\Rbb[{\rm km}] / d_{10})^2$, where $d_{10}$ is distance in units of 10 kpc.
The multicolour disk black body model {\sc diskbb} has an inner disc temperature $\Tdisc$ and normalization $\Kdisc\!=\!(\Rdisc[{\rm km}] / d_{10})^2 \cos i$, where $i$ is the inclination of the accretion disc.
An extension to the {\sc diskbb} model, the so called ``p-free'' model {\sc diskpbb}, 
is determined by the inner disc temperature $\Tpfree$, normalization $\Kpfree$ and the ``$p$-parameter'', 
that describes the radial dependence of the disc temperature as $T(r)\!\propto\!r^{-p}$ (see e.g. \citealt{WMM01}).
In the standard {\sc diskbb} model $p\!=\!0.75$, while $p \!\approx\!0.5$ for super-Eddington (slim) accretion discs with or without outflows (see \citealt{ACL88,WMM01,PLF07}).
We also used a Comptonization model {\sc comptt}, that has four parameters; the seed photon temperature for Comptonization $T_{\rm seed}$, electron temperature $T_{\rm e}$, optical depth $\tau$ and normalization $K_{\rm c}$.

The crude estimation of the spectral hardness for the 2010 data (using X-ray colours in the 0.3--2 keV and 2--10 keV bands) in Fig. \ref{fig:pn_lc_hr} 
indicates that the spectral shape remains constant during the observation even though the flux decays by $\sim$50 per cent.
As such short term flux variations of ULXs are not so commonly observed, we attempted to model this by splitting the observation into ``high'' and ``low'' flux parts as shown in Fig. \ref{fig:pn_lc_hr}.
However, the only confident conclusion from this analysis was that the spectral shape remains constant (see Table \ref{tab:bestfits} and Fig \ref{fig:spectra}a).
In all the models we used the parameter variations are so subtle that the parameter errors do not permit clear and significant conclusions to be made. 
We therefore used the time-averaged spectrum to study which phenomenological models provide the best fits to the spectral data.

\subsection{Strong spectral curvature in the 2010 data}

\begin{figure*}
\centerline{\epsfig{file=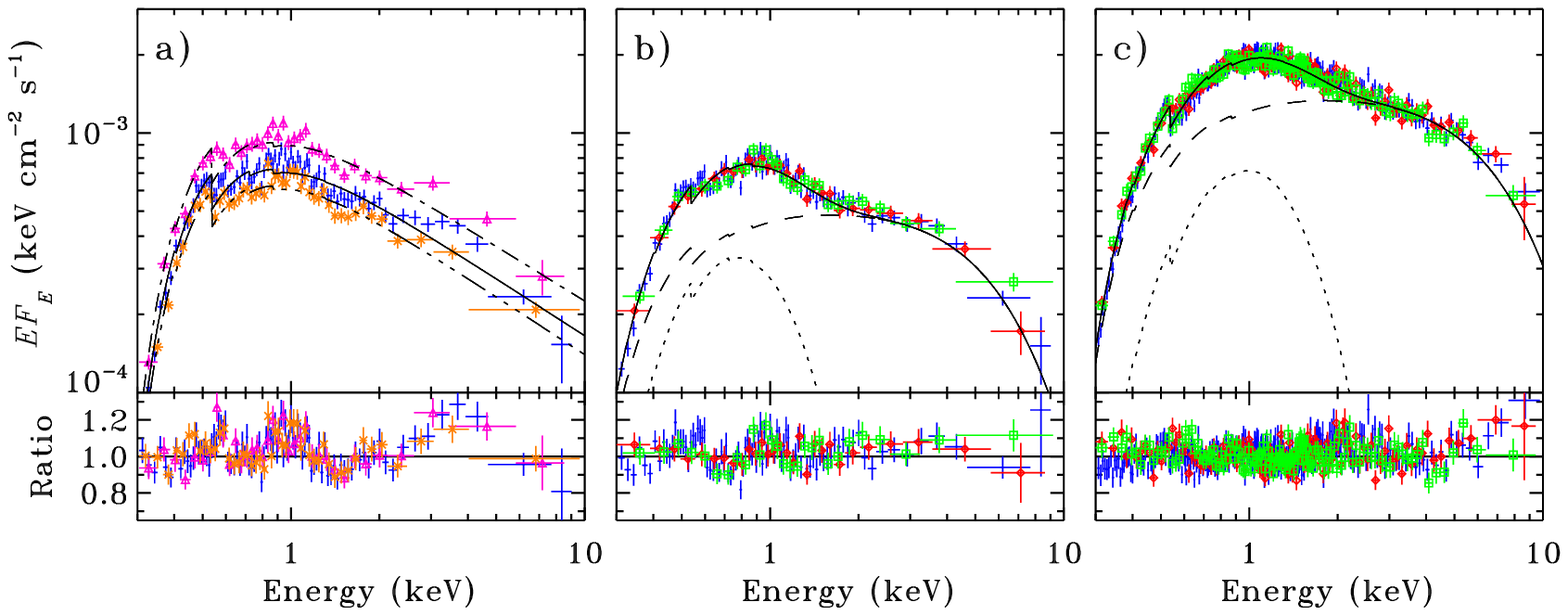,width=\textwidth}}
\caption{a) The high state (pink triangles), time-averaged (blue crosses) and low state (brown diagonal crosses) EPIC-pn $EF_E$ spectra for the 2010 data, modelled with an absorbed {\sc powerlaw}.
The dot-dashed, solid and three-dot-dashed lines show the model spectrum for high, time-averaged and low flux states, respectively.
The spectral residuals are very similar to the \swift/XRT low flux spectra \citep{GKF10} and the shape of the residuals indicate that the chosen continuum is incorrect.
It is also visibly by eye that there is a clear spectral curvature break at approximately 3--4 keV.
b) The time averaged $EF_E$ spectra of the 2010 observation, modelled with {\sc bb$+$diskpbb}; red diamonds and green squares denote EPIC-mos1 and mos2 spectra, respectively.
The dotted and dashed lines show the {\sc bb} and {\sc diskpbb} model components, respectively.
The spectral curvature/break can be described with this model.
c) The time averaged $EF_E$ spectra of the 2004 observation. 
In all panels the spectral data are binned for visual clarity.}
\label{fig:spectra}
\end{figure*}

We started our spectral analysis of the 2010 data with the simplest absorbed {\sc powerlaw} model, and found that the photon index $\Gamma\!\approx\!2.73\!\pm\!0.03$ (see Fig. \ref{fig:spectra}a).
This value is slightly lower than in the lower flux \xmm\ observation of 2002 \citep{DMG04,FK09,KP09}, and of the \swift/XRT monitoring \citep{GKF10}, where $\Gamma\!\approx\!3$.
The fit gave $\chired\!\approx\!1.16$ for $724$ degrees of freedom (d.o.f.) and -- as commonly observed in ULX spectra -- 
the fit could be improved by adding a cool thermal component to the model.
Adding a {\sc diskbb} component results to $\chired\!\approx\!1.10$, and similarly adding a {\sc bb} component yields $\chired\!\approx\!1.09$ for two d.o.f. less.
The obtained temperatures of $0.2$ for the {\sc diskbb} model and $0.16$ keV for the {\sc bb} model are typical for ULXs (see \citealt{WMR06,FK09,KP09}).
The fits, however, are not statistically acceptable (rejection probability $P_{\rm rej}\!>\!95$ per cent).
A closer look at the residuals (see Fig. \ref{fig:spectra}a) reveal a clear reason for this.

\begin{table*}
 \begin{minipage}{\linewidth}
  \centering
  \caption{Best-fitting parameters for the 0.3--10 keV data. In the {\sc diskpbb} model, the $p$-parameter was fixed to 0.5. 
	    In the {\sc tbabs $\times$ (bb $+$ diskbb)} model for the 2010 data, there is a degeneracy between the spectral components.
	    The {\sc bb} and {\sc diskbb} model components can be swapped around, and fits are statistically similar.
	    We assumed $\Tdisc = T_{\rm seed}$ in the {\sc diskbb $+$ comptt} model.
  }
  \label{tab:bestfits}
 \begin{tabular}{c c c c c c c c c c}
\hline \hline
state	 	& $\NH$ 			& $T_1$ 			& $K_1$ 		& $T_2$ or $T_{\rm e}$		& $K_2$ 			& $\Gamma$ or $\tau$		& $\Ecut$ 		& $K_{\rm \Gamma}$ 		& $\chi^2$ / d.o.f.  \\
		& ($10^{21}$ cm$^{-2}$)		& (keV)				&  			& (keV)				& [$\times10^{-3}$]		&				& (keV) 		& [$\times10^{-3}$]    		& \\
\hline 
\multicolumn{10}{c}{{\sc tbabs $\times$ powerlaw} } \\
2004 		& $1.84_{-0.03}^{+0.03}$	& $...$				& $...$			& $...$				& $...$				&  $2.634_{-0.013}^{+0.013}$ 	& $...$			& $2.56_{-0.03}^{+0.03}$	& $1654/1398$\\
high 		& $1.36_{-0.10}^{+0.11}$	& $...$				& $...$			& $...$				& $...$				&  $2.70_{-0.05}^{+0.05}$ 	& $...$			& $1.13_{-0.04}^{+0.04}$	& $513/471$\\
av. 		& $1.38_{-0.06}^{+0.06}$	& $...$				& $...$			& $...$				& $...$				&  $2.73_{-0.03}^{+0.03}$	& $...$			& $0.89_{-0.02}^{+0.02}$	& $840/724$\\
low 		& $1.40_{-0.09}^{+0.09}$	& $...$				& $...$			& $...$				& $...$				&  $2.74_{-0.03}^{+0.04}$	& $...$			& $0.77_{-0.02}^{+0.02}$	& $587/545$\\

\multicolumn{10}{c}{{\sc tbabs $\times$ (bb $+$ powerlaw)} } \\
2004 		& $1.54_{-0.07}^{+0.07}$	& $0.233_{-0.011}^{+0.0010}$	& $25_{-5}^{+6}$	& $...$				& $...$				&  $2.49_{-0.03}^{+0.03}$ 	& $...$			& $2.11_{-0.08}^{+0.08}$	& $1574/1396$\\
high 		& $1.2_{-0.2}^{+0.2}$		& $0.17_{-0.02}^{+0.02}$	& $90_{-40}^{+80}$	& $...$				& $...$				&  $2.49_{-0.09}^{+0.09}$	& $...$			& $0.90_{-0.09}^{+0.09}$	& $488/469$\\
av. 		& $1.22_{-0.10}^{+0.10}$	& $0.163_{-0.013}^{+0.013}$	& $70_{-20}^{+40}$	& $...$				& $...$				&  $2.55_{-0.06}^{+0.05}$	& $...$			& $0.74_{-0.04}^{+0.04}$	& $789/722$\\
low 		& $1.29_{-0.14}^{+0.14}$	& $0.16_{-0.03}^{+0.03}$	& $40_{-20}^{+60}$	& $...$				& $...$				&  $2.62_{-0.08}^{+0.08}$	& $...$			& $0.68_{-0.06}^{+0.06}$	& $577/543$\\

\multicolumn{10}{c}{{\sc tbabs $\times$ (diskbb $+$ powerlaw)} } \\
2004 		& $1.55_{-0.07}^{+0.07}$	& $0.33_{-0.02}^{+0.02}$	& $5.8_{-1.4}^{+1.7}$	& $...$				& $...$				&  $2.47_{-0.04}^{+0.04}$ 	& $...$			& $2.04_{-0.11}^{+0.11}$	& $1590/1396 $\\
high 		& $1.3_{-0.2}^{+0.2}$		& $0.21_{-0.03}^{+0.03}$	& $40_{-20}^{+50}$	& $...$				& $...$				&  $2.49_{-0.10}^{+0.10}$	& $...$			& $0.90_{-0.10}^{+0.10}$	& $491/469$\\
av. 		& $1.32_{-0.10}^{+0.10}$	& $0.20_{-0.02}^{+0.02}$	& $28_{-12}^{+21}$	& $...$				& $...$				&  $2.55_{-0.06}^{+0.06}$	& $...$			& $0.74_{-0.05}^{+0.05}$	& $795/722$\\
low 		& $1.35_{-0.14}^{+0.15}$	& $0.20_{-0.03}^{+0.03}$	& $16_{-11}^{+40}$	& $...$				& $...$				&  $2.63_{-0.09}^{+0.08}$	& $...$			& $0.68_{-0.07}^{+0.06}$	& $578/543$\\

\multicolumn{10}{c}{{\sc tbabs $\times$ (bb $+$ cutoffpl)} } \\
2004 		& $1.02_{-0.11}^{+0.11}$	& $0.217_{-0.006}^{+0.006}$	& $63_{-9}^{+10}$	& $...$				& $...$				& $1.94_{-0.12}^{+0.11}$ 	& $6.4_{-1.1}^{+1.5}$	& $1.91_{-0.09}^{+0.09}$	& $1487/1395$\\
high 		& $0.5_{-0.3}^{+0.3}$		& $0.175_{-0.015}^{+0.014}$	& $130_{-40}^{+50}$	& $...$				& $...$				& $1.4_{-0.8}^{+0.6}$ 		& $2.9_{-1.2}^{+3.4}$	& $0.79_{-0.13}^{+0.12}$	& $477/468$\\
av. 		& $0.6_{-0.2}^{+0.2}$		& $0.173_{-0.009}^{+0.009}$	& $100_{-20}^{+20}$	& $...$				& $...$				& $1.5_{-0.4}^{+0.3}$ 		& $3.3_{-0.9}^{+1.5}$	& $0.64_{-0.06}^{+0.06}$	& $755/721$\\
low 		& $0.8_{-0.3}^{+0.3}$		& $0.17_{-0.02}^{+0.02}$	& $70_{-30}^{+40}$	& $...$				& $...$				& $2.0_{-0.5}^{+0.4}$ 		& $5_{-2}^{+7}$		& $0.63_{-0.07}^{+0.07}$	& $568/542$\\

\multicolumn{10}{c}{{\sc tbabs $\times$ (diskbb $+$ cutoffpl)} } \\
2004 		& $0.97_{-0.10}^{+0.11}$	& $0.308_{-0.011}^{+0.010}$	& $18_{-3}^{+3}$	& $...$				& $...$				& $1.6_{-0.2}^{+0.2}$ 		& $4.6_{-0.8}^{+1.2}$	& $1.57_{-0.15}^{+0.15}$	& $1493/1395$\\
high 		& $0.7_{-0.2}^{+0.2}$		& $0.25_{-0.03}^{+0.03}$	& $40_{-14}^{+23}$	& $...$				& $...$				& $0.5_{-1.2}^{+0.9}$ 		& $1.8_{-0.7}^{+1.6}$	& $0.6_{-0.2}^{+0.2}$		& $474/468$\\
av. 		& $0.75_{-0.11}^{+0.13}$	& $0.24_{-0.02}^{+0.02}$	& $32_{-7}^{+10}$	& $...$				& $...$				& $0.9_{-0.6}^{+0.5}$ 		& $2.1_{-0.5}^{+0.9}$	& $0.52_{-0.09}^{+0.09}$	& $747/721$\\
low 		& $0.8_{-0.2}^{+0.3}$		& $0.23_{-0.03}^{+0.03}$	& $26_{-9}^{+16}$	& $...$				& $...$				& $1.4_{-0.9}^{+0.6}$ 		& $3.0_{-1.2}^{+3.2}$	& $0.54_{-0.14}^{+0.11}$	& $565/542$\\

\multicolumn{10}{c}{{\sc tbabs $\times$ (bb $+$ diskbb)}} \\
2004 		& $0.09_{-0.03}^{+0.04}$	& $0.241_{-0.003}^{+0.003}$	& $85_{-4}^{+5}$	& $1.32_{-0.03}^{+0.03}$	& $61_{-5}^{+6}$		& $...$ 			& $...$			& $...$	  			& $1830/1396$\\
high 		& $0.19_{-0.13}^{+0.14}$	& $0.192_{-0.009}^{+0.009}$	& $110_{-20}^{+30}$	& $1.14_{-0.06}^{+0.07}$	& $53_{-12}^{+15}$		& $...$				& $...$			& $...$	  			& $483/469$\\
av. 		& $0.15_{-0.07}^{+0.08}$	& $0.195_{-0.006}^{+0.005}$	& $84_{-10}^{+13}$	& $1.17_{-0.04}^{+0.04}$	& $38_{-5}^{+7}$		& $...$				& $...$			& $...$	  			& $785/722$\\
low 		& $<0.21$			& $0.198_{-0.009}^{+0.008}$	& $66_{-11}^{+16}$	& $1.13_{-0.06}^{+0.06}$	& $36_{-8}^{+10}$		& $...$				& $...$			& $...$	  			& $597/543$\\

\multicolumn{10}{c}{{\sc tbabs $\times$ (bb $+$ diskpbb)}} \\
2004 		& $0.87_{-0.03}^{+0.03}$	& $0.228_{-0.005}^{+0.005}$	& $64_{-5}^{+6}$	& $1.91_{-0.06}^{+0.07}$	& $3.9_{-0.5}^{+0.6}$		& $...$ 			& $...$			& $...$	  			& $1522/1396$\\
high 		& $0.73_{-0.12}^{+0.13}$	& $0.174_{-0.012}^{+0.012}$	& $120_{-40}^{+50}$	& $1.6_{-0.2}^{+0.2}$		& $3.4_{-1.3}^{+1.8}$		& $...$				& $...$			& $...$   			& $476/469$\\
av. 		& $0.70_{-0.07}^{+0.07}$	& $0.176_{-0.007}^{+0.007}$	& $90_{-20}^{+20}$	& $1.63_{-0.09}^{+0.10}$	& $2.7_{-0.6}^{+0.8}$		& $...$				& $...$			& $...$	  			& $750/722$\\
low 		& $0.67_{-0.10}^{+0.11}$	& $0.179_{-0.011}^{+0.011}$	& $70_{-20}^{+20}$	& $1.55_{-0.12}^{+0.14}$	& $2.9_{-0.9}^{+1.2}$		& $...$				& $...$			& $...$	  			& $567/543$\\

\multicolumn{10}{c}{{\sc tbabs $\times$ (diskbb $+$ comptt)}} \\
2004 		& $0.71_{-0.07}^{+0.05}$	& $0.21_{-0.02}^{+0.02}$	& $78_{-14}^{+14}$	& $2.4_{-0.4}^{+0.6}$		& $1.4_{-0.2}^{+0.2}$		& $5.0_{-0.8}^{+0.7}$ 		& 			& 	  			& $1486/1395$\\
high 		& $0.8_{-0.2}^{+0.2}$		& $0.22_{-0.04}^{+0.04}$	& $60_{-30}^{+60}$	& $1.1_{-0.2}^{+0.4}$		& $0.9_{-0.3}^{+0.3}$		& $10_{-3}^{+7}$		& 			&    				& $476/468$\\
av. 		& $0.75_{-0.10}^{+0.11}$	& $0.22_{-0.02}^{+0.02}$	& $50_{-20}^{+20}$	& $1.2_{-0.2}^{+0.2}$		& $0.75_{-0.11}^{+0.12}$	& $9_{-2}^{+2}$			& 			& 	  			& $749/721$\\
low 		& $0.71_{-0.15}^{+0.17}$	& $0.20_{-0.03}^{+0.04}$	& $50_{-20}^{+50}$	& $1.3_{-0.2}^{+0.4}$		& $0.7_{-0.2}^{+0.2}$		& $8_{-2}^{+2}$			& 			& 	  			& $565/542$\\

\hline 

\end{tabular}

\end{minipage}
\end{table*}

The general shape of the {\sc powerlaw} based model residuals show clear evidence of a spectral break around $4$ keV.
To account for these residuals, we replaced the {\sc powerlaw} component with the cutoff powerlaw model {\sc cutoffpl}.
As the residuals indicated, we find the best fitting e-folding energies at $\sim\!3$ keV, depending on the additional soft excess model.
These fits are statistically acceptable with $\chired\!\approx\!1.04$ for 721 d.o.f.

The photon indices also flatten when fitting the data with the {\sc bb$+$cutoffpl} (or {\sc diskbb$+$cutoffpl}) model.
As flat powerlaw spectra are produced with advective disc models and discs with outflows (see e.g. \citealt{WMM01,PLF07}), we replaced the cutoff powerlaw component with the {\sc diskpbb} model.
We fixed the $p$-parameter to $p\!=\!0.5$ to mimic an advective/outflowing disc.
We find that this model ({\sc bb$+$diskpbb}) fits the observed spectral data very well giving $\chired\!\approx\!1.04$ for 722 d.o.f. (see Fig. \ref{fig:spectra}b).
The best fitting inner disc temperature is $\Tpfree\!=\!1.63_{-0.09}^{+0.10}$ keV and the cool {\sc bb} component had $\Tbb\!=\!0.176\!\pm\!0.007$ keV.
The inner disc radius (obtained from the fit normalizations and assuming $i=0\degr$ to get the lower limit) of the {\sc diskpbb} model is roughly $\Rdisc\!\gtrsim\!20$ km, 
whereas the black body component radius is $\Rbb\!\sim\!3000$ km.
The values are consistent with a StMBH interpretation, 
if we associate the $\Rdisc$ with the innermost stable orbit and $\Rbb$ with the photosphere of the outflow at $\sim$100 Schwarzschild radii.
These radii, however, should be taken as order of magnitude estimates, because several factors -- like colour corrections and corrections related to inner disc boundary conditions -- can alter the results (see e.g. \citealt{GZP99,MKM00}). 
Advection ($p\!\approx\!0.5$) is strongly required by the data. 
If we set $p\!=\!0.75$ -- i.e. replace the {\sc diskpbb} component with the {\sc diskbb} model -- the fits become significantly worse (see Table \ref{tab:bestfits}).

We also fitted the data with a disc plus Comptonization model ({\sc diskbb + comptt}).
The {\sc comptt} model is commonly used in describing the spectral curvature in ULXs (see e.g. \citealt{GRD09,MSR11}).
We assumed that the seed photon temperature for Comptonization is the same as the inner disc temperature ($T_{\rm seed}\!=\!\Tdisc$).
A very low electron temperature $T_{\rm e}\!\approx\!1.2$ keV ``corona'' of moderate optical depth $\tau\!\sim\!10$ can describe the spectrum well ($\chired\!\approx\!1.04$ for 721 d.o.f.), 
given that the $\sim\!0.2$ keV temperature accretion disc dominates the emission below 1 keV.

\begin{table*}
  \centering
  \caption{Best-fitting parameters of powerlaw based models in the 2--10 keV band. The flux in this band $F_{2-10}$ is given in units of $10^{-12}\,\ergcm2s$ and the break energy $\Ebr$ in keV.
	  The second last column shows the change of $\chi^2$ value when the {\sc powerlaw} model is replaced by the {\sc bknpower} model and the last column is the probability of chance improvement from the F-test.}
  \label{tab:2to10keV}
 \begin{tabular}{c c c c c c c c c c}
\hline \hline
			& \multicolumn{3}{c}{\sc powerlaw}		& \multicolumn{4}{c}{\sc bknpower}& \\
OBSERVATION	 	& $\Gamma $ 	& $\chi^2$ / d.o.f. 	& $F_{2-10}$	& $\Gamma_1 $ 	& $\Ebr$ 	& $\Gamma_2$ 	&  $\chi^2$ / d.o.f. & $\Delta \chi^2$ 	& $P$ \\

\hline 
2004 data 		& $2.56 \pm 0.03$	& $837/828$	& $2.65^{+0.04}_{-0.05}$	& $2.51 \pm 0.04$	&  $5.4^{+0.4}_{-0.5}$	& $3.1 \pm 0.2$		& $817/826$	& $19.9$	& $5\!\times\!10^{-5}$\\
2010 data 		& $2.66 \pm 0.07$ 	& $177/193$	& $0.84^{+0.03}_{-0.04}$	& $2.34^{+0.12}_{-0.18}$&  $4.1^{+0.4}_{-0.7}$ 	& $3.4^{+0.3}_{-0.4}$	& $150/191$	& $27.0$	& $1\!\times\!10^{-7}$\\
\hline

\end{tabular}
\end{table*}

\subsection{Comparison with the 2004 data}

The 2004 high luminosity state data provides an interesting comparison to these results (see Fig. \ref{fig:spectra}c).
The overall spectral shape in the 2004 observation is similar to the 2010 observation, with one important exception.
We saw that the spectral cutoff in the 2004 data occurred at a higher energy.
This can be seen in all the models we used (see Table \ref{tab:bestfits}); 
in the {\sc cutoffpl} model the e-folding energy increases to $\sim$4--7 keV, or the inner disc temperature increases to $\sim\!2$ keV when we use the {\sc bb$+$diskpbb} model, 
or the electron temperature increases to $\sim\!2.4$ keV in the {\sc diskbb + comptt} model.
As these results depend on the chosen continuum model, we also made the spectral fits in the 2--10 keV band using the {\sc bknpower} model so that we can compare our results to those reported in \citet{SRW06} and \citet{GRD09}.
The best fitting spectral parameters in the 2004 data are consistent with these studies. 
We find that adding a break at $\Ebr\!=\!5.4^{+0.4}_{-0.5}$ keV improves the fit over a simple powerlaw model (see Table \ref{tab:2to10keV}).
In the 2010 low luminosity state data, we instead find the break at $\Ebr\!=\!4.1^{+0.4}_{-0.7}$ keV.
We can thus see that, at higher than 90 per cent confidence level, the spectral break in the 2010 data occurs at a lower energy than in the higher luminosity state \xmm\ observation of 2004.

Another interesting difference between the 2004 and 2010 spectra is the temperature increase of the cool thermal component with increasing luminosity.
However, this trend is highly model dependent, to a point where it all together disappears when using the {\sc diskbb + comptt} model.
If the trend is real, it is clearly different from several ULXs where an inverse relation between the temperature and luminosity has been detected \citep{FK07,KP09}.
The 2002 low/soft state observation of \Ho\ also deviates from these trends \citep{FK09,KP09,FS11}, which raises an interesting question.
If the cool thermal component is associated with the outflow, 
the correlation between temperature and the luminosity seems to contradict the model prediction of \citet{PLF07} developed for highly super-Eddington accretion rates.
This might indicate a more complicated accretion disc structure at a mildly super-Eddington state and that our current understanding of ULXs is far from complete. 
Also, it is very likely that the phenomenological models we use to model ULX spectra are oversimplified, 
and they do not describe the complex nature of super-Eddington accretion flows accurately enough. 
The fact that the temperature of the cool component depends so strongly on the other model components is a stark reminder 
that it is necessary to know the mechanisms that produce the harder emission above $\sim\!2$ keV before we can address the nature of the soft spectral component.

\section{Discussion}\label{sec:discussion}

In the 2010 \xmm\ observation of \Ho, we have detected a significant spectral break at $\Ebr\!\approx\!4$ keV.
By comparing this to an earlier \xmm\ observation taken in 2004, where a break has been detected at $\Ebr\!\approx\!5.4$ keV \citep{SRW06,GRD09}, we see that in this low flux state, the break occurs at a lower energy.
We found in our analysis that the spectral break might not be caused by a low temperature corona as was suggested by the previous studies \citep{SRW06,GRD09,MSR11}.
Instead, the spectrum -- and the spectral break -- can be modelled with a hot, advective slim disc model, with a contribution from an optically thick outflow radiating at energies below 2 keV.
The flux--$\Ebr$ correlation in \Ho\ can therefore be related to advection in the disc; the spectral break moves to higher energies as the mass accretion rate increases.
This behaviour is also seen when spectra of other ULXs are fitted with hot disc models; 
the inner disc temperatures seem to increase with increasing luminosities \citep{SRW06} and in fact even some ULX spectra can also be fitted with an advective disc model (including Holmberg IX X-1, \citealt{TKN06}, see below).
Furthermore, the central parts of advective super-Eddington accretion discs can ``overheat'' and the colour correction factor (and therefore the colour temperature) can be very large close to the black hole \citep{B98}.
Also, according to this model, the colour temperature is a strong function of the viscosity parameter $\alpha$ (\citealt{SS73}).
Small changes in $\alpha$ could, therefore, cause the colour temperature to increase so that the spectral curvature could even move beyond the spectral range of \xmm.
This mechanism could possibly explain the very high inner disc temperatures that have been detected in some ULXs \citep{GRD09}. 

It is interesting to note that the flux--$\Ebr$ trend seems to be different to the ULX in Holmberg IX, 
where \citet{VDR10} has reported opposite behaviour (see however \citealt{KYY10}, who come to a different conclusion using the same data).
\citet{VDR10} attributes the flux--$\Ebr$ anti-correlation in Holmberg IX ULX to increased accretion disc winds as the luminosity increases.
They argue that the wind increases material in the corona, 
which in their view causes the (unknown) acceleration mechanism to share the energy among an increasing amount of particles, leading to a smaller equilibrium electron temperature.
This mechanism could produce the trend of the break energy being anti-correlated with the observed flux, although clearly this trend does not always hold (see \citealt{VDR10,KYY10,MSR11}).
However, the detected flux--$\Ebr$ correlation in \Ho\ suggest that this mechanism is unlikely to be present in this particular case.
We can speculate that in Holmberg IX ULX the flux--$\Ebr$ anti-correlation might instead be related to the outflow.
If the outflows are intermittent \citep{O2007} -- such that they can be replaced by stronger advection and weaker outflows -- 
the hotter inner part of the accretion disc could either be visible or obscured, thus causing the spectral break to appear at different energies.
On the other hand, this opposite flux--$\Ebr$ trend might also be related to the accretion flow properties.
If small variations in the $\alpha$ parameter can be unrelated to changes in the mass accretion rate, then the colour correction factor can vary considerably independently of the luminosity.
Such variations can, in principle, also cause the spectral break to appear at different energies, and effectively eliminate any flux--$\Ebr$ trend.

What seems necessary is that we need at least three ingredients to explain the spectral properties of ULXs in the StMBH scenario; outflows/winds, advection and coronae above the disc.
Obviously, by permutating all these components in the spectral modelling we can explain the data, but real understanding of the processes require good understanding of the underlying trends.
Discovery of such trends requires, however, frequent monitoring programs of ULXs with deep \xmm\ exposures similar to \swift/XRT monitoring (\citealt{KF09,S09,GKF10}).
\xmm\ observations are needed especially to increase spectral sensitivity at energies closer to 10 keV, to detect the spectral curvature/breaks.
These trends (if they are confirmed) could then be used to compare with predictions from simulations of super-Eddington accretion discs \citep{O2007,OM11} to infer the accretion flow properties.
Alternatively, by analysing a large sample of high quality \xmm\ spectra of different ULXs, one might be able to infer these trends.
However, if different ULXs show different flux--$\Ebr$ trends (as these results seem to indicate), such attempt might not yield much information about these spectral features. 
Furthermore, if some ULXs are IMBHs while others are perhaps super-Eddington StMBHs any possible luminosity--$\Ebr$ trend (or the lack of it) might not be visible in the whole ULX population.

\section{Conclusions}

We have performed spectral analysis of an \xmm\ observation of \Ho\ taken in 2010 during its low/soft state.
The observed flux decayed roughly by $\sim$50 per cent during the observation.
We found that the spectral shape did not change significantly during the flux decay.
The spectrum itself was complex, showing a soft excess and a high energy curvature/break.
The spectral data could be best fitted with a model consisting of a black body plus a cutoff powerlaw model. 
The cutoff powerlaw model could also be replaced with a slim disc model. 

By comparing the spectral break energy $\Ebr$ to earlier \xmm\ data taken in 2004, we could determine that $\Ebr$ is correlated with the observed flux.
We note that this behaviour is opposite to that in Holmberg IX ULX, where anti-correlation between $\Ebr$ and flux has been claimed.
We speculate that the reason for the apparent difference could be that in \Ho\ the corona is replaced with a slim accretion disc, when the system is only mildly above the Eddington limit.
Alternatively, this difference might arise from different geometries of the outflow and the inner parts of the accretion discs in these two ULXs.
Clearly these results are tentative and to better understand the nature of the spectral curvature/breaks and its flux dependence additional data spanning a much larger bandpass are required.
Thus, repeated simultaneous observations with \xmm\ and the upcoming {\it NuStar} mission (with a combined bandpass of 0.3--80 keV) could help to clarify the differences and the observed trends.

\section*{Acknowledgments}

This work was supported by the Emil Aaltonen Foundation and the Finnish Graduate School in Astronomy and Space Physics (JJEK) and the Academy of Finland grant 127512 (JP). 
SAF is the recipient of an ARC Postdoctoral Fellowship, funded by grant DP110102889. SAF acknowledges funding from the UK STFC.
PK and FG acknowledge partial support from NASA grant NNX10AF86G.
This publication was based on observations obtained with \xmm, an ESA science mission with instruments and contributions directly funded by ESA Member States and NASA.
We thank Norbert Schartel and the staff of the \xmm\ Science Operation Centre for performing this Target of Opportunity observation.
We also thank the referee for helpful comments and suggestions.

\bibliographystyle{mn2e}

\label{lastpage}

\end{document}